\documentclass[12pt,twoside]{article}
\usepackage[latin1]{inputenc}
\usepackage{times}
\usepackage{epsfig}
\usepackage{dsfont}
\usepackage{amsfonts}
\usepackage{float}
\usepackage{amsmath}
\usepackage{latexsym}
\usepackage{xr} 
\usepackage[authoryear]{natbib}
\usepackage{color}
\usepackage{multirow}
\usepackage{makecell,url}

\relax 
\allowdisplaybreaks[3]

\numberwithin{equation}{section}

\newcommand{\blue}[1]{\textcolor{black}{#1}}

\begin{document}

\title{\Large \bf On the role of benchmarking data sets and simulations in method comparison studies}
\author{Sarah Friedrich$^{1, 2}$ and  Tim Friede$^{3,4}$ \\[1ex] 
{\small $^1$ Institute of Mathematics, University of Augsburg, Germany}\\
{\small $^2$ Centre for Advanced Analytics and Predictive Sciences (CAAPS), University of Augsburg, Germany}\\
{\small $^3$ Department of Medical Statistics, University Medical Center Göttingen, Germany}\\
{\small $^4$ DZHK (German Centre for Cardiovascular Research), Partner Site Göttingen, Göttingen, Germany} \\
{\small sarah.friedrich@uni-a.de, tim.friede@med.uni-goettingen.de}
}

\maketitle

%
%

\begin{abstract}
	Method comparisons are essential to provide recommendations and guidance for applied researchers, who often have to choose from a plethora of available approaches. While many comparisons exist in the literature, these are often not neutral but favour a novel method. Apart from the choice of design and a proper reporting of the findings, there are different approaches concerning the underlying data for such method comparison studies.
	Most manuscripts on statistical methodology rely on simulation studies and provide a single real-world data set as an example to motivate and illustrate the methodology investigated. In the context of supervised learning, in contrast, methods are often evaluated using so-called benchmarking data sets, i.e.~real-world data that serve as gold standard in the community. Simulation studies, on the other hand, are much less common in this context. The aim of this paper is to investigate differences and similarities between these approaches, to discuss their advantages and disadvantages and ultimately to develop new approaches to the evaluation of methods picking the best of both worlds. To this aim, we borrow ideas from different contexts such as mixed methods research and Clinical Scenario Evaluation.  
\end{abstract}

\noindent{\bf Keywords:} Benchmarking; Machine Learning; Neutral Comparison Studies; Simulation studies.

\vfill
\vfill

\newpage

\section{Introduction}
The process of examining a research question empirically consists of several steps ranging from study design and data analysis to the interpretation of the results \citep{Friedrich2021}. Each of these steps involves decisions to be made: Which trial design is adequate for answering the research question? Which analysis methods are available for the kind of data collected and what has to be taken into account when interpreting the results? The steps of this process have also been discussed in the context of drug development using so-called Clinical Scenario Evaluation (CSE) \citep{Benda2010,Friede2010,dmitrienko2017clinical}. The CSE framework consists of three core elements: options, assumptions and metrics. The different options for each step are compared using the respective metrics and taking the underlying assumptions into account. 
Simulation studies can be used in different stages of this process: to determine an adequate design including, for example, sample size planning\blue{, to inform a subsequent trial of `ideal' parameter settings or expected outcomes (\emph{in silico} clinical trials)} as well as to compare different methods for the statistical analysis, see also \cite{Morris2019} for an overview.
A relevant aspect in this context is to distinguish between models and methods. As \cite{Morris2019} put it: ``The term `method' is generic. Most often it refers to a model for analysis, but might refer to a design or some procedure
(such as a decision rule).'' In this sense, the method comprises questions such as `How to fit a model?' and `How to draw inference?'. It is important to keep this in mind when considering the comparison of different methods.

In order to choose the \blue{`best'} approach for a specific design or data analysis, fair comparisons between existing methods are essential. \blue{One can argue that a comparison study will never be completely neutral or fair in practice. In this paper, we therefore adopt the definition of `neutral comparisons' given by \cite{Boulesteix2013}, namely that the focus of the article should be on the comparison itself instead of introducing a novel method, that the authors should be reasonably neutral and that the study should be designed and evaluated in a rational way. See also \cite{Strobl2022} for a similar discussion.}
Recently, it has been noted in the context of data analysis that there is a tendency to over-optimistic reporting of the performance of new methods and a lack of neutral comparison studies in the literature, see e.g.~\cite{Boulesteix2015,Boulesteix2013,Boulesteix2017,van2018benchmarking,Weber2019,buchka2021optimistic,Niessl2021,Pawel2022}. \blue{For example, \cite{Boulesteix2013} found only 12 comparison studies out of 55 articles on supervised classification in a literature search.}
Neutral comparison studies, however, are essential to guarantee a fair comparison of existing methods across different scenarios, thus allowing an applied researcher to determine the \blue{`best'} method for her or his situation. Similar criticism can also be formulated for the case when simulations are used in a trial design context.
When planning a comparison study, a lot of options exists, see e.g.~\cite{Niessl2021} for an overview of design and analysis options. Beside the choice of an adequate design and proper reporting of the results, however, the question arises on what kind of data the methods should be compared. Here, different disciplines have different approaches.

When publishing a paper on statistical methodology, manuscripts usually consist of three major parts: theoretical derivations revealing (often asymptotic) properties of the proposed method, a simulation study investigating the small sample behavior and/or comparing the proposed method to relevant competitors, and a data example demonstrating the application of the proposed method to real-world data. 
Biometrical Journal, for example, explicitly encourages authors to ``include a description of the problem and a section detailing the application of the new methodology to the problem''\footnote{\url{https://onlinelibrary.wiley.com/page/journal/15214036/homepage/productinformation.html}}.
In this `classical' format, the simulation study usually covers a wide range of scenarios, while the application to real-world data is often restricted to a single example data set. Depending on the kind of paper, i.e. focusing on data analysis or on trial designs, this data example can serve different roles. For example, \cite{Muetze2020} use a data set on pediatric multiple sclerosis to demonstrate how this study could have been stopped early, if different monitoring procedures had been used. On the other hand, they also discover scenarios where the observed follow-up time does not provide enough information yet.

In the context of machine learning (ML), particularly supervised learning, another approach is common: the performance of methods is usually compared on so-called benchmark data sets, which serve as gold standard and enable comparison of methods on real-world data. That way, they serve as an important step in the process between method development and clinical use \citep{Friedrich2021ai}. Simulation studies, on the other hand, are much less common in \blue{most} ML applications. \blue{In some areas of ML, however, simulations also play a role, for example as digital twins \citep{Batty2018}. This idea is currently employed in different application areas, ranging from industrial applications \citep{Jiang2021} to agriculture \citep{Pylianidis2021} and precision medicine \citep{Voigt2021}.
	Another exemption is learning from simulated data, see \cite{Michoel2007,Gecgel2019,Behboodi2019} for some examples in different application areas.}

\blue{Sometimes, methods are also compared on several real data sets as well as on synthetic data, see \cite{Hothorn2005} and \cite{Bischl2013} for early examples.} Recently, a number of so-called \emph{empirical studies} have been published in statistical papers, see e.g.~\cite{Stegherr2021,Wiksten2016,Seide2019,Turner2021} for examples. These papers demonstrate the method(s) of interest on a variety of real-world data sets, thus not solely relying on simulations. 

When comparing these approaches, matters are complicated by the fact that different terms are used in different application areas. While most papers in the context of bioinformatics, machine learning and artificial intelligence (AI) talk of benchmarking \citep[e.g.][]{Koch2021,buchka2021optimistic,raji2021ai,Dwivedi2020}, the terms `empirical study' \citep{Stegherr2021,Seide2019}, `empirical evaluation' \citep{Turner2021} or `empirical comparison' \citep{Wiksten2016} are also common. \cite{Clark2022}, on the other hand, mention neither benchmarking nor empirical study, but describe their approach as ``[$\dots$]  a separate and novel contribution to the assessment on the model classes [$\dots$] by a pairwise assessment on the population of networks that the research community would choose to fit them on.'' 
On the other hand, the term benchmark is also used to refer to an `ideal' method, for example an approach that has complete information which would not be available in an actual trial \citep{Mozgunov2022}. 
This makes systematic reviews of the literature difficult and to the best of our knowledge, no systematic comparison of the different approaches exists to date.


In this paper, we aim to investigate differences and similarities between the approaches, discuss their advantages and disadvantages and develop a new framework aimed at picking \blue{`the best of both worlds'}. Furthermore, we 
identify tasks that are necessary to be addressed by the scientific community in order to enable the combination of both approaches on a regular basis.

The paper is organized as follows: In Section \ref{sec:methods} we give more formal definitions of the concepts of benchmarking and simulation studies and contrast the pros and cons of the two approaches. We summarize our findings in some recommendations in Section \ref{sec:recommendations} and use these to critically discuss some examples in Section \ref{sec:examples}. We close with a discussion in Section \ref{sec:discussion}.

\section{Differentiating Benchmarking and Simulation Studies}\label{sec:methods}

\subsection{Simulation studies} \label{sec:simu}
Simulation studies are a common tool in statistics and complement theoretical derivations of statistical methods. The basic idea is to investigate the behaviour of a method when applied to synthetic data, i.~e.~data with known properties \citep{boulesteix2020introduction}. In particular, simulation studies can serve different purposes:
(i) Compare several existing methods to determine which performs \blue{`best'} in a given scenario, (ii) investigate small sample properties of a method in addition to asymptotic results based on theory, (iii) study the robustness of a method if underlying assumptions are violated and (iv) support assessments of complex design scenarios and sample size planning \blue{for an individual study or a series of experiments. Aspect (iv) is}
especially relevant in the context of 
complex study designs such as adaptive designs \citep{Friede2011, Friede2020} as well as in the Clinical Scenario Evaluation framework \citep{Benda2010} \blue{and is conceptually different from the other approaches: Instead of drawing conclusions for a wide range of applications or settings, the focus here is on one specific study (or a series of studies) and the aim is to find the `best' design for this specific application.} 
It is worth mentioning that simulation studies often reflect a frequentist approach, where the true  parameters are fixed but unknown values. In Bayesian statistics, simulation studies are mainly used to analyse frequentist properties of posterior-based decision rules \citep{morita2010evaluating,thall1994practical}. Similarly, it is more common to modify the choice of priors in sensitivity analyses \citep[e.g.~Chapter 6 of][]{gelman1995bayesian}.
Thus, simulation studies can be viewed as a model-based approach in the sense that mathematical concepts and models need to be known (or assumed to be known). Based on these models, we can investigate which data we can fit them to and where their limits are.

Recently, \cite{Morris2019} have shown, that simulation studies are often poorly designed, analyzed and reported. To overcome this issue, they provide recommendations and guidelines for the design, implementation and reporting of simulation studies. \blue{Earlier, \cite{Burton2006} already proposed implementing a protocol for simulation studies and provided a checklist of important considerations for the design of such a study.} A detailed explanation aimed at applied researchers is given by \cite{boulesteix2020introduction}. \blue{\cite{Chipman2022} suggest to employ methods from design and analysis of experiments when planning and conducting simulation studies, e.g.~factorial designs and ANOVA methods.}
Following these guidelines can improve the conduct of simulation studies and provide a basis for fair and neutral comparisons based on simulated data.

\subsection{Benchmarking} \label{sec:benchmarking}
Benchmarking \blue{originates} from computer sciences \blue{and is a relatively new concept} \citep{raji2021ai}. According to \cite{Xie2021}, a benchmark study is a ``systematic comparison between computational methods, in which all of them are applied to a gold standard dataset and the success of their [$\dots$] predictions are summarized in terms of quantitative metrics [$\dots$]''. \blue{As \cite{Hothorn2005} describe it, benchmarking aims at ``[measuring] performances in a landscape of learning algorithms''. The assessment of an algorithm's quality by means of , e.g.~cross-validation, started in the 1970's with the pioneering work of \cite{Stone1974}. Later, the focus shifted to comparisons of algorithms rather than performance assessment tasks and benchmarking algorithms on various data sets came up in the 1990's with a ``shift from rationalism to empiricism'' \citep{Church2019}. Since then, b}enchmarking has \blue{established a} tradition in machine learning, specifically in the context of supervised learning, where competitions 
such as ImageNet \citep{deng2009imagenet} fostered the comparison of different methods  on a common data set. \blue{This trend increased in the last few years due to greater availability of open data. For example,} the Neural Information Processing Systems conference (\url{https://neurips.cc/}) introduced a new track specifically for data sets and benchmarks in 2021 \citep{neurips}.
\blue{D}ata repositories such as the UC Irvine Machine Learning Repository \citep{uci} or Kaggle (\url{https://www.kaggle.com/}) provide platforms for benchmarking data sets.
Such platforms also exist for specific applications. For ECGs, for example, methodological development has been hampered until the recent publication of the PTB-XL data set, which is hosted by PhysioNet  \citep{physionet,Strodthoff2021}.

In contrast to simulation studies, benchmarking provides a data-driven approach. This approach might often be closer to the questions faced by applied users of statistical models: Given my research question and my data, which is the \blue{`best'} approach I can choose for an adequate analysis? 
Moreover, for algorithmic approaches without an underlying mathematical model as in many AI applications, where the focus is primarily on prediction instead of inference, designing a simulation study is not straightforward.
In supervised learning, where the data are equipped with labels and  thus allow for calculating performance measures, benchmarking provides a comparison between methods that is close to real-world applications. In the context of unsupervised learning, the situation is more involved, since the data do not contain known labels. 
\blue{In the special case of cluster analysis, for example, researchers often use data sets with known labels to evaluate their algorithms, although the true labels are actually unknown in clustering applications \citep{Ullmann2022}.
	Thus, the role of the test data is not as clear as it is in the supervised learning context \citep{Ullmann2021} and the choice of performance evaluation methods is more complex. This also enables jumping to over-optimistic conclusions in cluster analysis, as demonstrated by \cite{Ullmann2022} on both synthetic and real data. \cite{van2018benchmarking} provide guidelines for benchmarking in cluster analysis and point out the importance}
of repositories equipped with metadata to provide a good data basis. This issue is also discussed by \cite{Zimmermann2019}, who criticises a lack of suitable data in unsupervised learning. 
To reduce the issue of unknown class membership in real data, \cite{van2018benchmarking} recommend to combine ``simulations and empirical data as these may yield complementary information''. \cite{Zimmermann2019} also advocates the creation of artificial data as an alternative to real-world data, but stresses the need of realistic data-generating mechanisms, that capture the important properties of real data.

A range of papers discuss issues with over-optimistic benchmarking studies and provide guidelines on the conduct and reporting of fair comparison studies, see \cite{Weber2019,Kreutz2019,Zimmermann2019,van2018benchmarking,Niessl2021,buchka2021optimistic} and the references cited therein.
A particularly important aspect are the data resources, i.e.~availability of relevant real-world data for a given problem. In the context of network analysis, \cite{Clark2022} approach the problem of representativity by choosing a population of networks based on publications in a premier journal for social network analyses. This population of networks has successfully completed peer-review and can thus be ``deemed of sufficient scientific interest'' \citep{Clark2022}. 
\blue{In most research areas, however, there is still no gold standard of benchmark data sets. This can lead to other major problems like data leakage, i.e.~spurious findings that arise as artefacts of the data collection process or pre-processing steps. \cite{Kapoor2022} show that this is a wide-spread issue and leads to severe reproducibility failures in many different research areas.
	Another aspect of over-optimistic reporting is that comparisons are usually not based on sound statistical test decisions, see \cite{Hothorn2005} as well as \cite{boulesteix2015statistical} for a hypothesis testing framework. This also relates to sample size calculations:}
From a statistical perspective, the benchmarking data sets serve as `cases'. Thus, it is important to compare the methods on a sufficient number of cases. This can be calculated in advance using methods of sample size calculation \citep{Boulesteix2015}.

\subsection{Comparison}\label{sec:comparison}

As described above, benchmarking and simulation studies provide two different approaches to a similar problem: evaluating the performance of several alternative methods based on data, simulated or real. While simulation studies present a more theoretic approach, where the underlying statistical model and some theoretical concepts of the data-generating process have to be known, benchmarking provides a data-driven approach. We will now compare the two approaches with regard to their respective advantages and disadvantages. A short summary of the findings is presented in Table~\ref{tab:vergleich}. 

A huge advantage of simulation studies is that the `ground truth' is known, although sometimes it cannot be derived analytically but is assessed through simulations \citep{Austin2010}. Thus, it is possible to accurately investigate proposed methods with respect to, e.g., bias, coverage probability or control of the type-I error, since the underlying true values are known by design. Another advantage is that (practically) as much data as required can be simulated, if sufficient computer power is available. Thus, in contrast to real-world data sets, there are hardly any restrictions on sample size.
On the other hand, simulated data might not adequately reflect properties of real-world data and the generalizablity of the results may be limited. This aspect particularly comes into play when studying AI applications, which are often applied to complex, high-dimensional data sets. Adequately capturing the properties of this kind of data, particularly with respect to correlations and interactions between variables, might be difficult in a simulation study. 
Furthermore, the use of simulations can be limited by computational costs: For computationally expensive methods (e.g. Bayesian approaches) it might not be feasible to conduct a simulation, which requires a large number of simulation runs, in a reasonable amount of time.
Moreover, there exists an infinite space of possible parameters and simulation settings. Thus, a simulation study can only ever cover a tiny part of that space.
\blue{This makes the choice of the data-generating process highly subjective and concerns about the relevancy and plausibility of simulated data for real-world applications are warranted \citep{Boulesteix2017}. Thus,}
the settings should be chosen reasonably and interpreted with caution \citep{Boulesteix2015,Pawel2022}.

Benchmarking, on the other hand, is applied to real-world data and therefore allows an assessment of whether the choice of methods matters in practice.  Especially in the context of supervised learning, where the data themselves contain the `truth' and when the focus is on prediction rather than inference, benchmarking is closer to reality than a simulation study. 
\blue{Interestingly, many popular benchmarking data sets originally stem from the statistical literature, see \cite{Hothorn2005} for some examples.}
As \cite{Clark2022} call it, this approach takes a \emph{data-centric} viewpoint as opposed to a model-based viewpoint.
In other situations, however, it might matter that the ground truth is not known in a real-world data set, e.g.~in the context of hypothesis testing. 
Recently, benchmarking has been criticized for a number of reasons. As \cite{raji2021ai} point out, benchmarking data sets often fail to achieve the goal of `generality' they are imagined to possess. Instead, they are ``inherently specific, finite and contextual'' \citep{raji2021ai}. A central issue in this context is validity: how well does the data and the associated evaluation metric represent the given task? Are the questions investigated actually relevant to applicants in the field? Does the benchmark study represent relevant real-world data? See also the discussions in \cite{raji2021ai,Koch2021} and \cite{buchka2021optimistic} as well as \cite{bao2021s} for a practical example. A circumstance adding to this issue is that benchmarking studies often use samples of convenience, i.e.~data sets most easily accessible for the researchers \citep{Koch2021,raji2021ai}. These might either be widely spread data sets or chosen from a familiar context of the researchers \citep{buchka2021optimistic}. In the latter case, the results might not easily generalize to other situations. 
As \cite{Koch2021} observe, there is an increasing concentration on fewer and fewer data sets used in benchmarking over time and these have been introduced by just a handful of institutions. 
Thus, they might not even be neutral but potentially influenced by some objective or even sponsored by a specific firm or institution.
As a consequence, benchmarking data sets often possess a poor representation of real-world data. For example, many data sets used for training algorithms in natural language processing are only available in English and the majority of images on ImageNet stem from Western Countries \citep{raji2021ai}, thus potentially introducing bias in the machine learning algorithms. 
\blue{The major point of criticism with respect to over-optimistic benchmarking thus stems from the underlying data, with issues such as representativity, validity and data leakage. However, as \cite{Kapoor2022} point out, there are also other reasons for over-optimistic findings, such as choosing an evaluation metric that is not ideally suited for the task at hand.
}

\begin{table}[h]
	\centering
	\caption{Strengths and weaknesses of benchmarking and simulation studies. The $\checkmark$ should be interpreted as `has a tendency to perform better in this respect' rather than an absolute assessment of suitability.}
	\label{tab:vergleich}
	\begin{tabular}{c|c|c}
		& \textbf{Simulation Studies} & \textbf{Benchmarking}  \\ \hline
		Ground truth known & $\checkmark$ & $\times$ \\ 
		Unlimited data available & $\checkmark$ & $\times$\\
		Computational cost & $\times$ & $\checkmark$\\
		Closer to reality & $\times$ & $\checkmark$ \\
		Data-centric viewpoint & $\times$ & $\checkmark$ \\
		Model-based viewpoint & $\checkmark$ & $\times$ \\ 
		Used beyond method comparison &  $\checkmark$ & $\times$ \\ 
		Applicable to algorithmic approaches  & $\times$ & $\checkmark$ \\ 
	\end{tabular}
	
\end{table}

To further compare the approaches, we take the point of view of Clinical Scenario Evaluation (CSE).
The CSE framework consists of three core elements: options, assumptions and metrics \citep{Benda2010}. The metrics serve as tools for comparing the different options (which can be varied by the researcher) given the underlying assumptions (which are fixed but unknown). Some of the assumptions might be informed by previous studies whereas others have to rely on subject-matter knowledge only. Given the uncertainty, it is good practice to vary the assumptions in the sense of, e.g., sensitivity analyses. 
When viewing benchmarking and simulations in this framework, we get the following: For benchmarking, the competing `options' include the choice of comparators, the tuning of hyperparameters as well as the choice of statistical software. The `assumptions' in this setting are the data sets. Similar to CSE, they should span a range of optimistic, realistic and pessimistic situations. Based on these, the competing options can be compared using various metrics, which depend on the specific situation.

For a simulation study, the competing `options' consist of the various choices related to the simulation setting (simulation scenarios, choice of software, number of simulation runs, etc) as well as the choice of competitors. The `assumptions' here are those on the underlying data-generating process, i.e.~the mathematical or statistical model that provides the backbone of the simulation study. \blue{The important difference in the `assumptions' between benchmarking and simulation studies is that in the simulation study, the data-generating process is known to and chosen by the researcher. In benchmarking, on the other hand, we only observe a realization of the (unknown, underlying) data-generating process. While there exists a choice with respect to which data sets are being analyzed, the true data-generating process will always remain unknown. However, it is usually not necessary for the data analysis to completely know the data-generating process. In a nonparametric approach, for example, the assumptions on the underlying data distribution are usually rather weak. In a simulation study, in contrast, it is hardly possible not to specify the data-generating process in detail, even if fewer or less stringent assumptions are made in the analysis. For example, \cite{Muetze2016} evaluate permutation approaches under a variety of parametric distributions, which differ with respect to e.g.~skewness, see Table III in \cite{Muetze2016}. Similarly, \cite{Friedrich2017wild} investigate their wild bootstrap approach for nonparametric data in a simulation study where data are generated according to a variety of parametric distributions, reflecting both ordinal and continuous data settings.} The options can again be compared by a variety of metrics, which depend on the specific situation. An overview of these aspects is provided in Table \ref{tab:cse}. Table \ref{tab:metrics} provides an overview of different metrics used in simulation and benchmarking for a variety of statistical tasks. As we can see, there is no difference between benchmarking and simulations with respect to prediction or classification, since the true labels are contained in the data. With regard to the other tasks, all metrics applied to benchmarking data sets can also be applied to simulated data. Additionally, simulated data allows to compute metrics on the population level, such as bias or type-I error, which cannot be computed on a real data set.

\begin{table}[h]
	\centering
	\caption{Benchmarking and simulations in light of the CSE framework.}
	\label{tab:cse}
	\begin{tabular}{c|ccc}
		&  \textbf{Assumptions} & \textbf{Options} & \textbf{Metrics} \\ \hline
		Benchmarking   & Data sets & \makecell{Hyperparameters, \\ comparators,\\ choice of software} & \makecell{Performance measures\\ for specific situation} \\ \hline
		Simulation & Data-generating process & \makecell{Simulation setting, \\ comparators, \\ \blue{hyperparameters}, \\ \blue{choice of software}} & \makecell{Performance measures\\ for specific situation} \\
	\end{tabular}
\end{table}

\begin{table}[h]
	\centering
	\caption{Exemplary metrics for different statistical tasks used in benchmarking and simulation. The table is adapted from \cite{Morris2019}.\\
		CI = confidence interval, SE = standard error, AIC = Akaike Information Criterion, BIC = Bayesian Information Criterion, MSE = mean-squared error}
	\label{tab:metrics}
	\begin{tabular}{c|c|c}
		\textbf{Statistical Task}  & \textbf{Benchmarking} &  \textbf{Simulation}\\
		\hline
		Estimation &  \makecell{empirical SE,  length of CI} &  \makecell{Bias, SE, MSE,\\ coverage,  length of CI}\\ \hline
		
		Hypothesis testing & -- & \makecell{Type-I error, power}\\ \hline
		
		Model selection & AIC/BIC & \makecell{Sensitivity/ specificity \\for covariate selection, \\ AIC/BIC} \\ \hline
		
		Prediction/ Classification & \makecell{\blue{Predictive accuracy, i.e.} calibration \\ \blue{and} discrimination}&  \makecell{\blue{Predictive accuracy, i.e.} calibration \\ \blue{and} discrimination} \\
		\hline
		Study design& -- & \makecell{Sample size, duration, \\ power/precision}\\
	\end{tabular}
	
\end{table}

Finally, it should be noted that in both simulation studies and benchmarking, the choice of the comparators, an adequate study design and transparent reporting of results and limitations are fundamental. In particular, the chosen `methods' need to be clearly defined, including possible pre-processing steps or parameter tuning, and the latter must be optimized for each method separately \citep{van2018benchmarking,Weber2019}.

\section{Recommendations}\label{sec:recommendations}
Our discussion shows that each approach has its merits and shortcomings. In light of these considerations, we recommend to \blue{take} a broader point of view and learn from the other discipline, respectively. Thus, we encourage statisticians to perform benchmarking analyses additionally to the traditional simulation study, where possible and feasible. As seen previously, the latter is not always the case: Applying benchmarking in trial design studies, for example, is only feasible if the proposed design results in a shorter observation period compared to the design underlying the data. See \cite{Muetze2020} for an example, where the required target information was not achieved in the data example based on the proposed methods. On the other hand, simulation studies might not be feasible, if the data and models involved are computationally very expensive. Examples include resampling approaches in complex models \citep{ditzhaus2018more}, model-based recursive partitioning \citep{Huber2021} as well as many machine learning methods such as boosting \citep{Thomas2017,Klinkhammer2022}. Particularly when combining these approaches, computation times and/or memory issues may become problematic.

It should be noted that this recommendation stretches beyond encouraging the individual user to apply both benchmarking and simulations in his or her next study. In order to adequately address this issue, the whole community is needed. In particular, the necessary infrastructure needs to be established. This starts with providing and extending databases and data repositories that enable large-scale benchmarking studies. To address the issues raised in the context of benchmarking, these data need to be adequately curated, equipped with meta data and cautiously monitored \citep{Zimmermann2019,Koch2021,raji2021ai,van2018benchmarking,Strodthoff2021}. 
In addition, the community needs to establish what can be viewed as a `gold standard data set' for a given application. In this context, there is a role to play for the scientific societies in developing guidelines and providing recommendations. \blue{Moreover, this notion also extends to scientific journals. Here, publication bias and fear of rejection still provide pressure for publishing `new' and `better' approaches. Moreover, most high-ranking statistical journals do not mention comparison studies in their scopes \citep{Boulesteix2017}. Thus, to enable more comparison studies, the way they are evaluated by journal editors and reviewers needs to change as well.}

To the best of our knowledge, a sensible approach for combining the two worlds is missing. Ideas for bridging the gap could be taken from mixed methods research \citep{creswell2017designing,hesse2010mixed}, where quantitative and qualitative research methods are combined in order to maximize the strengths and minimize the weaknesses of each type of data. An overview \blue{of the approaches in mixed methods research and how they translate to our situation} is given in Table \ref{tab:mmr}.
Integration, i.e. the interaction between the different components of the study, is an essential aspect in mixed methods research \citep{o2010three}. According to \cite{creswell2011best}, there are three core approaches to integrate different forms of data: merging data, connecting data and embedding data. 
In the context of simulation studies and benchmarking, `embedding data' can be viewed as the current practice in many statistical manuscripts, where the results of a simulation study (large, primary design) are enhanced by additionally analyzing a data example (secondary priority). 
Similarly, the recent trend towards combining so-called empirical studies based on several data sets with simulated data can be viewed as merging data: The two types of data (simulated and real) are analyzed separately and the results are combined in a discussion section. The third idea in mixed methods research is connecting data. The aim here is to use the results obtained from one type of data (e.g.~qualitative data) to inform a subsequent study (e.g.~by developing new items for a quantitative data collection). The order of the two types of data may be reversed here. 
\blue{In the context of simulation studies and benchmarking, several existing approaches fall into this category: 
	\begin{enumerate}
		\item Simulating data based on a real data example: Many simulation studies aim to mimic a real data example, see e.g.~\cite{friedrich2017nonparametric,Bluhmki2018,Ohneberg2019,friedrich2020causal,graf2022} to name just a few. Note, however, that the degree to which the simulated and real data overlap, varies greatly: Sometimes simulations are simply using the estimated mean and (co-)variances from the real data, sometimes other aspects such as length of follow-up or number of observed events are simulated based on observed data.
		Moreover, new simulation approaches are often inspired by a data example, for which the existing methods are not adequate. \cite{Sylvestre2008}, for example, developed a simulation approach for time-dependent covariates based on a permutation approach, see also \cite{permalgo}. Similarly,  \cite{crowther2013} extended existing approaches for simulating time-to-event data motivated by a data example for which the existing simulation approaches seemed too simplistic.
		\item Reconstructing data sets based on published information: To counter-act the problem of available individual patient data, several approaches have been considered to reconstruct these data based on published information. The method by \cite{Guyot2012}, for example, allows for reconstruction of survival data based on published Kaplan-Meier-curves. Some examples for method comparisons based on this approach include \cite{Royston2019,Dormuth2022}. A similar approach has been proposed by \cite{Bluhmki2019}, who use resampling approaches based on published Nelson-Aalen plots for simulating realistic data.
		\item Plasmode simulation studies: Another relevant concept in this context are Plasmode Simulations, see e.g.~\cite{Franklin2014}. Here, the idea is to use part of a real-world data set (for example to capture difficult relationships between large numbers of covariates) and to artificially create an outcome (e.g.~treatment effect) of the researcher's choice.
	\end{enumerate}
	As mentioned above, it is possible to reverse the order of data types in mixed methods research. This approach might also be possible in our context, even though it has, to our knowledge, not been implemented yet.
	Speaking in terms of clinical trials, simulation studies represent prospective, experimental designs, but are conducted under `laboratory conditions'. Benchmarking studies, on the other hand, are usually conducted retrospectively, i.e.~based on already existing data. However, in some situations it is possible to use information obtained from simulations to inform subsequent trials:
	\begin{enumerate}
		\item Simulation studies can be used to explore settings that are especially relevant for the method under consideration. Afterwards, the method can be verified in benchmark data sets which represent the settings identified by the simulations. This approach is comparable to several existing approaches in different fields. For example, simulation is increasingly used to determine a promising set of input parameters for a biomanufacturing system, for example the production of antibodies for drug development \citep{Wang2019}. Another example are \emph{in silico} clinical trials, i.e.~``trials for pharmacological therapies or medical devices based on modelling and simulation technologies'' \citep{Musuamba2021}.   
		Here, the idea is to use individualized simulations to speed up the process of drug development by informing clinical trials beforehand on expected outcomes and possible modifications \citep{Viceconti2016}. \emph{In silico} clinical trials can thus either complement or replace \emph{in vivo} clinical trials. An early example for the application of \emph{in silico} clinical trials is given by \cite{Clermont2004}, who investigate the feasibility of this approach in clinical trials of severe sepsis.
		\item Comparison of design approaches: Our second suggestion relates to the aspect of trial designs. As stated above, this is a field that is not yet present in benchmarking experiments, although simulation studies allow to investigate trial designs as well. Thus, one could envisage comparing several design approaches (which proved promising in simulations) in the real world. More specifically, one would conduct, for instance, a prospective, randomized controlled trial, where the `interventions' are different trial designs.
		One idea in this direction are so-called SWATs (Study Within A Trial) \citep{Clark2022SWAT,swat}. To date, SWATs are mainly used to evaluate the effectiveness of e.g.~recruitment strategies, but could potentially be extended to cover more complex design aspects as well. In some situations, this type of evaluation might be unrealistic, since heterogeneity between studies would be too large to ensure comparable results. Similarly, conducting several clinical trials would often be too time-consuming. Thus, experiments where this approach might be possible would have to be more homogeneous and/or less time-consuming. One example is the context of economics, where experiments are sometimes conducted as business simulation games \citep{Jobjoernsson2022}.    
	\end{enumerate}
}

%

\begin{table}[ht]
	\centering
	\caption{Comparison of data integration approaches in mixed methods research and in the context of simulation and benchmarking.}
	\label{tab:mmr}
	\begin{tabular}{c|c|c|c}
		\makecell{\textbf{Approaches for} \\ \textbf{integration}} & \makecell{\textbf{Mixed methods} \\ \textbf{research}} & \makecell{\textbf{Simulation and}\\ \textbf{Benchmarking}} & \textbf{Examples} \\ \hline
		Merging data & \makecell{Combine qualitative data \\
			(e.g.~texts or images)\\
			with quantitative data} & \makecell{Combination of \\ empirical study on \\ several data sets\\ and simulation} & \makecell{\cite{Seide2019}: \\ Empirical study on 40\\ data sets complemented \\ by simulation study} 
		\\ \hline
		
		Connecting data & \makecell{Use information from \\ one data analysis \\
			(quantitative or qualitative) \\ to inform a subsequent \\data collection \\ (qualitative or quantitative) }& \makecell{Simulation study \\ inspired  by data \\ example;\\ \blue{Plasmode Simulations;}\\ \blue{Reconstruction of data sets;}\\ \blue{Using simulation results to}\\ \blue{inform subsequent studies}} & \makecell{\cite{friedrich2020causal}: \\ Simulation study inspired \\ by COVID-19 data; \\ \blue{\cite{Franklin2014}:}\\ \blue{Plasmode simulations;}\\ \blue{\cite{Dormuth2022}:}\\ \blue{Reconstruction of data sets}}\\ \hline
		
		Embedding data & \makecell{Data set of secondary \\ priority is embedded within \\ a larger, primary design} & \makecell{Simulation study with \\ additional analysis of\\ a data example} & \makecell{\cite{friedrich2017permuting}:\\ Extensive simulations\\ complemented by one \\ exemplary data analysis}\\
	\end{tabular}
\end{table}

\section{Examples}\label{sec:examples}
\blue{In the following,}
we discuss some exemplary studies with regard to possible improvements. We deliberately chose studies, in which at least one of the authors was involved, since the purpose is not to criticize others but to discuss pros and cons of existing studies in light of the arguments made in this paper.

\begin{figure}
	\centering
	\includegraphics[width=0.5\textwidth]{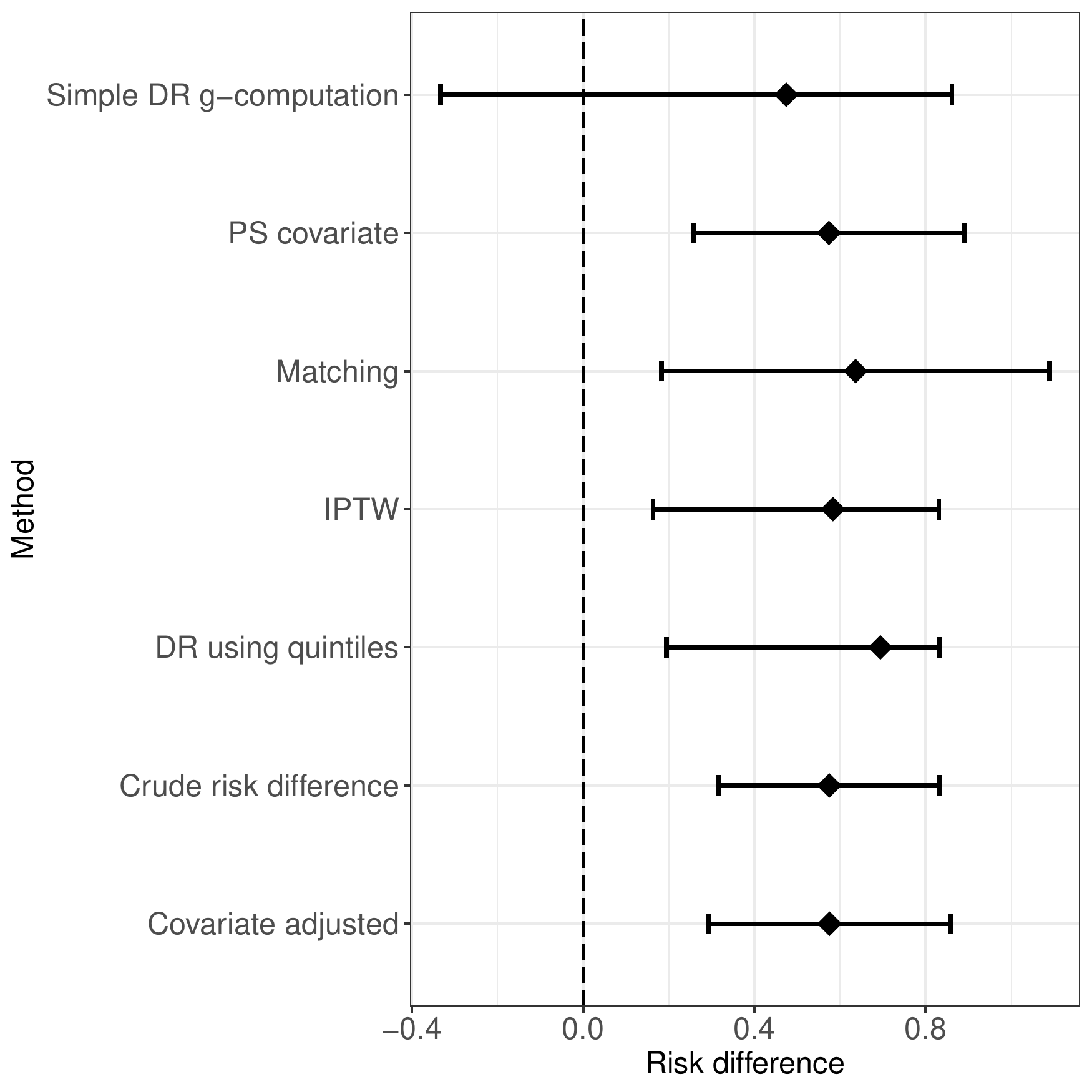}
	\caption{Estimated risk differences with 95\% confidence
		intervals obtained by the different methods. For details on the different methods, see \cite{friedrich2020causal}.
	}
	\label{fig:causal_example}
\end{figure}

\paragraph{A simulation study}
Motivated by an early non-randomized trial in COVID-19, \cite{friedrich2020causal} investigate the behavior of different causal inference methods in a large simulation study. In terms of Table~\ref{tab:mmr}, data are \emph{connected} in this manuscript. The study can be considered neutral, since no new methodology is proposed and neither of the authors have been involved in developing any of the approaches under consideration. The parameters underlying some simulation scenarios are motivated by the data example, while other scenarios were taken from another paper \citep{Austin2007}. However, the authors did not accurately follow the recommended ADEMP structure by \cite{Morris2019} or the CSE framework \cite{Benda2010}.
An important aspect to note here is that even though the data are artificial, the ground truth is not known in all scenarios. In particular, the `true' causal risk difference is estimated based on simulated counterfactual outcomes in large data sets ($n = 10,000$) and the underlying parameters are iteratively modified, until the desired risk difference is approximately reached \citep{Austin2010}. Based on these values, the methods are compared with respect to bias, length of confidence intervals, root mean squared error and coverage probability, since the statistical task is estimation (cf.~Table~\ref{tab:metrics}).
Although the paper is motivated by a real data example, the authors did not include an analysis of this data example in the final manuscript. This was due to the fact that the data set failed to illustrate the methods compared. In particular, some methods investigated in the simulations could not be applied to the data example and all methods essentially came to the same conclusion, see Figure \ref{fig:causal_example}. This was due to the major statistical and design issues in the original study that could not be rectified by more elaborate analysis methods. However, this case demonstrates that picking a simple data example can result in misleading conclusions and should thus be avoided. 
To sum up, \cite{friedrich2020causal} provide an example of a thorough simulation study, but without comparing the methods on real data. Thus, it remains unclear whether the theoretical results observed in the simulations would lead to different conclusions in real-world applications.

\paragraph{An empirical study}
An example of an empirical study is \cite{Stegherr2021}. Here, estimators typically used in the study of adverse events with varying follow-up times are compared in 17 randomized clinical trials. 
The properties of these estimators have been analyzed and discussed previously. In particular, a special issue in \textit{Pharmaceutical Statistics} was dedicated to the topic \citep{specialissue}. There, methods were demonstrated on single data examples \citep{proctor2016,bender2016,allignol2016}. Moreover, \cite{Stegherr2021a} compared the methods on artificial data in a simulation study.

The aim of this empirical study is to ``investigate and demonstrate which biases can occur in practice''. Data collection, inclusion criteria, analyses methods and the set-up of the meta-analysis are explained in \cite{Stegherr2020a}.
The methods are compared to a gold-standard method by investigating the ratios of the probability estimates obtained with the different estimators divided by the probability estimates obtained with the gold-standard. Thus, the ground truth is assumed to be obtained through the gold-standard method in this example, which has implications for the assessment of some properties such as bias.

Due to the opportunistic sample of data sets used in the empirical study, however, generalizability is limited \citep{Stegherr2021}. In particular, more than two thirds of the trials included in the study stem from oncology and adverse events were heterogeneous due to their backgrounds in different therapeutic areas. Thus, the sample is not representative of clinical studies in general.
In order to improve this study and in light of the recommendations above, a database of randomized trials with time-to-event outcomes investigating adverse events would be needed. This, however, brings along issues of data protection, which were addressed in the study by analyzing the data at the respective sponsor's site and only transferring aggregated results, i.e. the calculations are done in a distributed fashion.

\paragraph{An empirical study complemented by simulations}
As a final example, we consider \cite{Seide2019}. This empirical study on 40 meta-analyses is complemented by a simulation study, and is thus in line with our advice to combine both approaches. According to Table\ref{tab:mmr}, this study thus \emph{merges} empirical and artificial data.

In particular, an empirical data set of 40 meta-analyses was extracted from recently published reviews in a systematic manner. Similar to \cite{Stegherr2021}, the different methods were compared to a gold-standard approach and the ratios of the obtained point estimates were considered as metrics. Moreover, the length of the empirical confidence intervals was compared on the empirical data. In the simulation study, coverage probabilities could additionally be used as metric, since the ground truth was known in this case. 
As the authors state ``A consideration of all meta-analyses might have led to a more complete picture, but was not feasible with the resources of this project [$\dots$]'', highlighting again the need for adequate databases such as Cochrane Database of Systematic Reviews.

\section{Discussion}\label{sec:discussion}

Method comparison studies are an important tool to provide recommendations for \blue{both applied and methodological} researchers. \blue{While applied researchers wonder about the `best' method to pick for their data analysis, method comparisons can also help methodological researchers in determining potential for further improvements or identifying limitations of existing methods and thus a need for the development of new approaches.}
In order to yield valid results, however, \blue{method comparison studies} need to be conducted in a neutral fashion, not biased towards novel methods.
\blue{In practice, one might argue that a comparison study can never be entirely neutral \citep{Strobl2022}, nor can a benchmarking analysis, since in any case there are choices to make (regarding the underlying data-generating process or the data sets). Thus, the term `neutral' in this paper should be interpreted as: ``being [$\dots$] focused on the comparison of existing methods already described elsewhere rather than on a new prototype method being introduced [$\dots$]'' \citep{Strobl2022}.
}
In this paper, we have focused on the aspect of the underlying data: these could be real data sets from practical applications or artificial data.
The idea for these considerations was born while working on a white paper of the German Consortium in Statistics (DAGStat, \url{www.dagstat.de}) on Artificial Intelligence \citep{Friedrich2021}. 
In this paper, we have introduced the two approaches and discussed their respective advantages and shortcomings. Since no approach is always superior to the other, we recommend to use a combination wherever possible \blue{and we have made some suggestions on how that could be achieved.} 

Some final remarks are in place. First, we have not discussed possible approaches to combining the results obtained on several data sets (real-world or artificial) to come to a final conclusion regarding the `best' method. Here, several approaches exist. Most commonly, the methods are ranked according to their performance and results are presented as summaries of this ranking, see \cite{Niessl2021} for a detailed discussion. As pointed out by \cite{Boulesteix2013}, the concepts of meta-analysis could also be extended for the framework of method comparison studies.
\blue{In this context, it should also be noted that answers like `method A performs universally better than method B'
	can not be expected from comparison studies. Instead, one should rather consider which aspects of the underlying data (real or simulated) are associated with the good or bad performance of a method, see \cite{Strobl2022} for an extensive discussion of this topic.}
Second, one of the major selling points for simulation studies is that the ground truth is usually known for artificial data. Although this is true in many applications, it cannot always be achieved. In the context of causal inference, for example, the `truth', i.e.~the true causal effect, is often estimated even in simulations \citep{Austin2010,friedrich2020causal}. The advantage of simulated data is, of course, that very large data sets can be generated on which to estimate the causal risk difference, for example, but this should be kept in mind.
\blue{Third, as mentioned briefly above, an important aspect in benchmarking studies is the availability of relevant real-world data. Two aspects need to be considered in this context: (i) Availability: The recent push for open science and, as a consequence, data sharing will hopefully continue to improve the availability of data sets, thus enabling more large-scale benchmarking studies. In particular, many journals now require or encourage data sharing and platforms like the UCI Machine Learning Repository \citep{uci}, Kaggle (\url{https://www.kaggle.com/}) and the NIH Data Sharing Repositories \citep{nihrepository} as well as the \texttt{R}-package \textbf{OpenML} \citep{openml} provide lots of data sets for benchmarking tasks.
	(ii) It is also important that data is in a standard format and of sufficient quality to make benchmarking possible. This includes, for example, the collection of meta data and a cautious monitoring of the data quality. As already noted in Section \ref{sec:benchmarking}, missing standards for the underlying data can lead to major problems fuelling the reproducibility crisis, such as data leakage \citep{Kapoor2022}.}

\noindent \textbf{Acknowledgement}
Support by the German Research Foundation DFG (grants FR 4121/2-1, FR 3070/3-1, FR 3070/4-1) is gratefully acknowledged.
\vspace*{1pc}

\noindent {\bf{Conflict of Interest}}

\noindent {\it{The authors have declared no conflict of interest. }}\\

\noindent {\bf{Data availability statement}}

\noindent {\it{The COVID-19 data used in Section \ref{sec:examples} is provided in Supplementary Table 1 of \cite{Gautret2020}.}

	\bibliographystyle{apalike}
	\bibliography{benchmarking}

\end{document}